\documentclass[aip,pra,reprint,amsfonts,amsmath,amssymb,superscriptaddress,10pt,floatfix]{revtex4-1}

\usepackage{etex}
\reserveinserts{28}

\usepackage[english]{babel}
\usepackage[utf8]{inputenc}

\usepackage{amsmath}
\usepackage{amsfonts}
\usepackage{amssymb}
\usepackage{amsthm}
\usepackage{mathtools}
\usepackage{graphicx}
\usepackage{xfrac}
\usepackage{nicefrac}

\usepackage{color}
\usepackage{array}
\usepackage{longtable}
\usepackage{calc}
\usepackage{multirow}
\usepackage{hhline}
\usepackage{ifthen}

\usepackage[version=4]{mhchem}

\usepackage[colorlinks]{hyperref}

\graphicspath{{./figures/}}

\hypersetup{citecolor=blue, urlcolor=blue, linkcolor=blue, colorlinks=true}

\begin{document}
\title{Solid--solid phase equilibria in the NaCl--KCl system}
\author{Jamshed Anwar}
\thanks{J. Anwar and C. Leitold contributed equally to this work.}
\affiliation{Department of Chemistry, Lancaster University, Lancaster, LA1 4YW, United Kingdom}
\author{Christian Leitold}
\thanks{J. Anwar and C. Leitold contributed equally to this work.}
\affiliation{Department of Chemical and Biomolecular Engineering, University of
Illinois at Urbana-Champaign, Urbana, IL 61801, USA}
\affiliation{Faculty of Physics, University of Vienna, 1090 Wien, Austria (present affiliation)}
\author{Baron Peters}
\affiliation{Department of Chemical and Biomolecular Engineering, University of
Illinois at Urbana-Champaign, Urbana, IL 61801, USA}
\affiliation{Department of Chemistry, University of Illinois at Urbana-Champaign, Urbana, IL 61801, USA}
\date{\today}
\begin{abstract}
Solid solutions, structurally ordered but compositionally disordered mixtures,
can form for salts, metals, and even organic compounds.  
The NaCl--KCl system forms a solid solution at all compositions between
657°C and 505°C. Below a critical temperature of 505°C, the system
exhibits a miscibility gap with coexisting Na-rich and K-rich rocksalt
phases. We calculate the phase diagram in this region using the semi-grand
canonical Widom method, which averages over virtual particle transmutations.
We verify our results by comparison with free energies calculated
from thermodynamic integration and extrapolate the location of the
critical point. The calculations reproduce the experimental phase diagram
remarkably well and illustrate how solid-solid equilibria and chemical
potentials, including those at metastable conditions, can be computed for
materials that form solid solutions.  
\end{abstract}
\maketitle

\section{Introduction}

Solid solutions are ubiquitous in metallurgy,\cite{Porter1992,kurz1986fundamentals}
in geochemistry,\cite{putnis_1992,Becker2006} in biomineralization,\cite{Addadi2003,Becker2005,Chave1952}
and in many other areas of modern materials science.\cite{Graef2007,Mikkelsen1982,Joshi2008,Androulakis2007,Eitel2002,George2019}
The atoms in a solid solution reside at regular lattice positions,
but the components are randomly intermixed on the lattice.
At sufficiently low temperatures, many solid solutions separate into periodic
phases with different compositions. Thus solid solutions are nearly
perfect realizations of the idealized lattice models that are widely
used to study phase transitions in statistical mechanics.\cite{ChandlerGreenBook,stanley1987introduction}
At low temperatures, some solid solutions form “coherent precipitates”,
i.e. solute-rich precipitates embedded in the surrounding solvent
matrix, with both phases sharing one unbroken lattice. Further, certain solid
solutions can be quenched to form enormous populations of nanoscale
coherent precipitates yielding materials with extraordinary mechanical,
magnetic, and heat transfer properties. Examples include precipitate-hardened
Ni--Ti--Al superalloys,\cite{khachaturyan2008theory,Pollock1994} Heusler
or half-Heusler magnetic materials,\cite{Chai2013} and radiation
resistant alloys for nuclear reactor claddings.\cite{Odette2000,Miller2003} 

For these materials (and for solids in general), the earliest stages
of crystallization, namely, nucleation and subsequent growth, are a
major determinant of the structure and hence the properties of the
resulting product. Consequently, fundamental studies of nucleation
and growth are essential, but challenging both
experimentally and theoretically.\cite{FaradayDisc2015,Vekilov2010,AnwarZahn2011}
Notable studies of nucleation and growth in solid solutions have employed
Kinetic Monte Carlo simulations.\cite{Wu2016,Zhang2016,Clouet2006,Marian2011,Morgan2018} However,  
rare events like nucleation still pose challenges for kinetic Monte Carlo and molecular simulation. The available empirical potentials for multicomponent systems rarely make accurate property predictions while \textit{ab initio} calculations are too costly to capture the statistical ensembles---a sampling issue.\cite{Curtarolo2005}  

\begin{figure}
\centering \includegraphics[width=0.5\columnwidth]{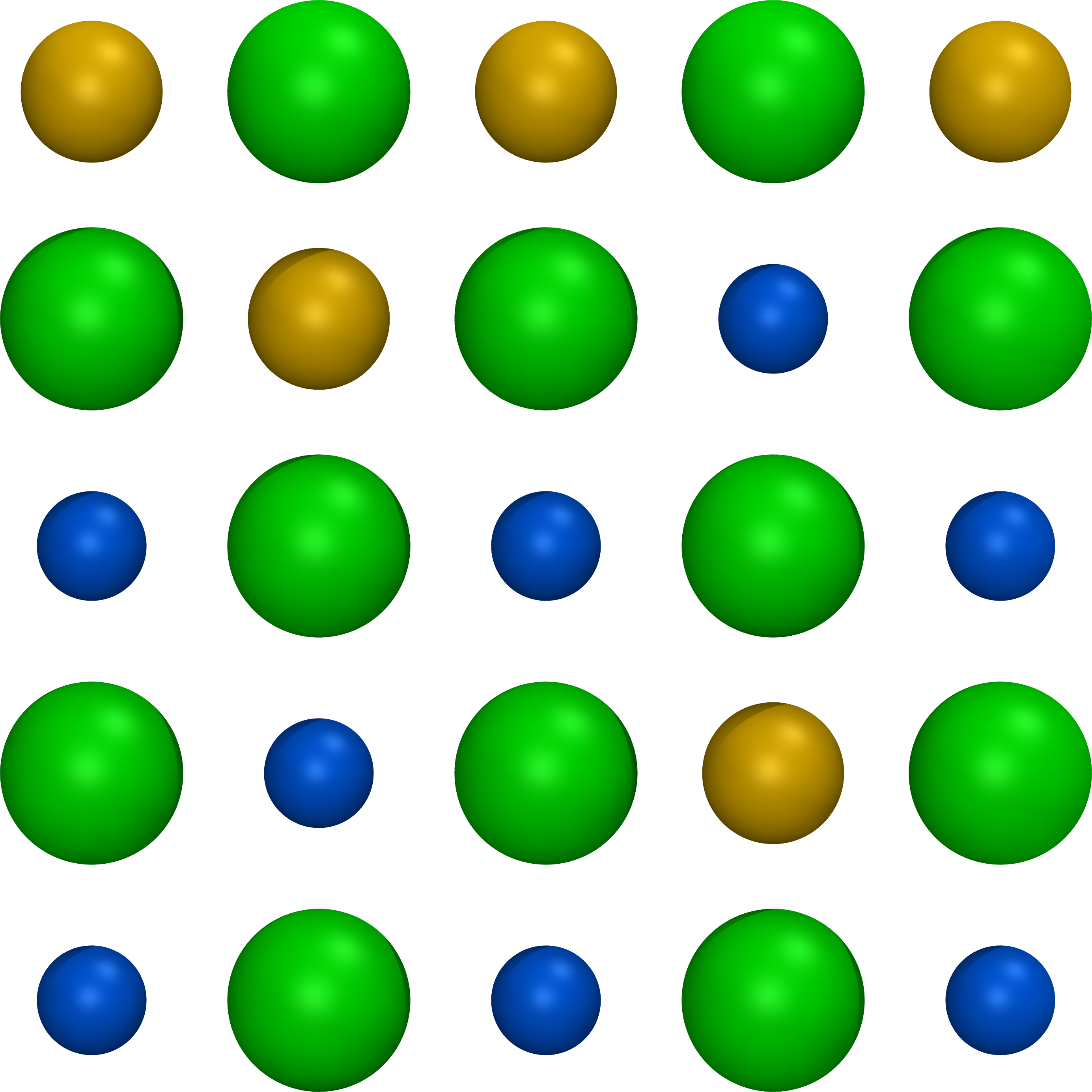}
\caption{Rendering of a single crystal plane in the NaCl--KCl solid solution.
While Cl (green) occupies a regular lattice, Na (blue) and K (yellow)
are randomly distributed on the remaining lattice sites. For clarity,
thermal fluctuations are not shown.}
\label{fig:rendering_perfect_lattice} 
\end{figure}

Simple abstract model systems often yield the most useful and generalizable
insights,\cite{Athenes2000,LeSarBook2013} but it is difficult to
quantitatively test their predictions against experiment. This paper
investigates a simple but a real model system, a binary mixture of
KCl and NaCl salts. Most molten salt mixtures exhibit a sharp eutectic
point, without the rather wide and unusual solid solution region seen
for NaCl--KCl. In the case of the NaCl--KCl solid solution, the chlorine
atoms occupy every other site of a simple cubic lattice, just like
in regular rocksalt. The sodium and potassium atoms, which occupy
the other half of the rocksalt lattice, are randomly distributed.
An illustration of the situation is shown in Fig.\,\ref{fig:rendering_perfect_lattice}.

\begin{figure}
\centering \includegraphics[width=1\columnwidth]{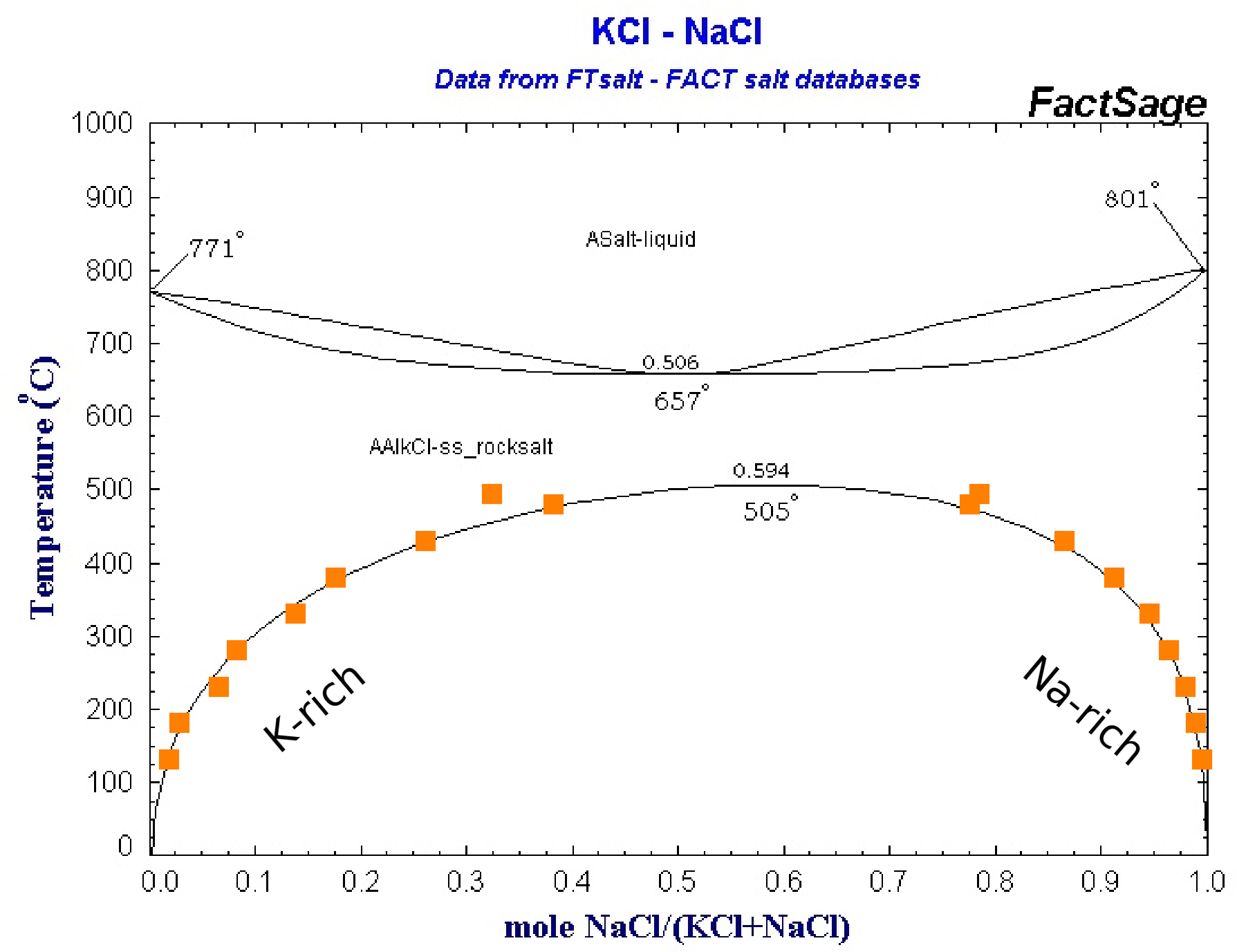}
\caption{Phase diagram for the NaCl--KCl system (obtained from the FTsalt database\cite{Bale2002}), which combines multiple experimental results along with the predicted data (orange squares) from the current simulation study.}
\label{fig:ftsalt} 
\end{figure}

The full phase diagram of the NaCl--KCl system, obtained from the
FTsalt database\cite{Bale2002} that combines multiple experimental
results (along with the predicted data from the current simulation study) is shown in Fig.\,\ref{fig:ftsalt}. The data displayed
in the phase diagram was originally compiled by JANAF\cite{Chase1985}
and has just recently been re-verified experimentally.\cite{Sergeev2015}
Of particular interest to our study is the fact that for low temperatures,
there exists a miscibility gap separating two compositionally different
solid solutions, an Na-rich and a K-rich one.

Here we have predicted the Na-rich / K-rich phase envelope in the
solid part of the phase diagram using molecular simulation. The calculated
phase diagram reproduces the experimental phase diagram remarkably
well (Fig.\,\ref{fig:ftsalt}). The calculations enable us
to predict the difference in chemical potential for NaCl and KCl
for any system composition, including the unstable region inside the
two-phase envelope. 

An important significance of the study is that
having an accurate phase diagram now enables {\it
quantitative} theoretical studies of nucleation to be carried out
for this model, as the chemical potential is now accurately defined
rather than assumed being equal to experimental data. Theoretical models
invariably show divergence from experimental data and the driving
force or chemical potential estimated from experimental data can be
markedly different from that of the model, introducing significant
uncertainty in the calculated nucleation rates.

In predicting the phase diagram we have employed two distinct approaches:
(i) the semi-grand canonical approach with the dependent variable
being composition as a function of set chemical potential difference,
and (ii) Widom's semi-grand canonical approach involving virtual
transmutation of one species to another to determine the chemical
potential difference (the dependent variable in this case) as a function
of set composition. Whilst we have demonstrated that both approaches
give consistent results, the Widom's semi-grand approach has significant
advantages and hence was our method of choice. We introduce the computational
model in Section\,\ref{sec:model}, outline the regular semi-grand
canonical ensemble in Section\,\ref{sec:semi-grand} and compare it with semi-grand canonical Widom method
in Section\,\ref{sec:semigrandwidom}. Details of the coexistence calculations are presented in Section\,\ref{sec:gibbscalculation}
and \ref{sec:phasediagcalculation}. The results are presented and discussed in
Section\,\ref{sec:results_discussion} followed by concluding remarks in Section\,\ref{sec:conclusions}.

\section{Model and methods}

\label{sec:methods}

\subsection{Force field}

\label{sec:model}

In order to model the NaCl--KCl system, we employ a Fumi--Tosi\cite{FumiTosi1964}
style force field. The full potential energy is given as a sum of
pair terms: 
\begin{equation}
U=\sum_{i<j}u(r_{ij}),
\end{equation}
where 
\begin{equation}
u(r_{ij})=A_{ij}\exp\left[B_{ij}(\sigma_{ij}-r_{ij})\right]-\frac{C_{ij}}{r_{ij}^{6}}-\frac{D_{ij}}{r_{ij}^{8}}+\frac{1}{4\pi\varepsilon_{0}}\frac{q_{i}q_{j}}{r_{ij}},
\end{equation}
and $r_{ij}$ is the distance between particles $i$ and $j$. The
parameters $A_{ij}$, $B_{ij}$, $C_{ij}$, $D_{ij}$, and $\sigma_{ij}$
are specific for the combination of elements of particles $i$ and
$j$ and hence there are six sets of parameters for all the possible
element pairings. These parameters, shown in Table\,\ref{tab:parameters},
are based on those given by Adams and MacDonald.\cite{Adams1974}

The final term in the force field is the electrostatic interaction,
where the charges (in units of the elementary charge q$_{\text{e}}$)
are +1 on Na and K and -1 on Cl. With the system of units employed
in this study, $\left(4\pi\varepsilon_{0}\right)^{-1}=1389.35$\,kJ/mol\,Å\,q$_{\text{e}}^{-2}$,
where $\varepsilon_{0}$ is the permitivity of free space.


Interestingly, in the parameter set tabulated by Adams and MacDonald,\cite{Adams1974}
the Cl--Cl $C$ parameter for Cl in NaCl (6985.70\,Å$^{6}$\,kJ\,mol$^{-1}$)
is different from that for Cl in KCl (7497.60\,Å$^{6}$\,kJ\,mol$^{-1}$).
Here we have utilised the average of these two parameter values. The same
is true for the Cl--Cl $B$ and $D$ parameters utilised in the present study.
For the Na--K hetero-interaction, the $\sigma$ parameter was taken as
the arithmetic mean of the homo-interactions. For the $C$ and $D$
parameters, the commonly employed geometric mean mixing rule was found
to be completely inappropriate: for instance, the Na--Cl $C$ parameter obtained from the geometric mean of
the Cl--Cl (in NaCl) and Na--Na $C$ parameters yielded a value
that was almost 25\% larger than the specified Na--Cl $C$ parameter
tabulated by Adams and MacDonald.\cite{Adams1974} On searching the
literature we came across the alternative mixing rule proposed by
Thakkar,\cite{Thakkar1984}

\begin{equation}
C_{ij}=\frac{2\alpha_{i}\alpha_{j}C_{i}C_{j}}{\alpha_{i}^{2}C_{i}+\alpha_{j}^{2}C_{j}},
\end{equation}

where $\alpha_{i}$ are (dimensionless) polarisabilities. Thakkar's
rule reproduced the Na--Cl $C$ and $D$ interaction parameters using
polarisabilities taken from Mayer\cite{Mayer1933} with no more than
4\% deviation or better. Consequently the Na--K $C$ and $D$ parameters
were estimated with Thakkar's mixing rule using the polarisabilities
of $\alpha_{\text{Na}}$=0.1820 and $\alpha_{\text{K}}$=0.8443 from
Mayer.\cite{Mayer1933}

\begin{table}[htbp]
\centering 
\global\long\def\arraystretch{1.3}%
\begin{tabular}{l|r r r r r}
	\hline \hline
	&        $A$&	$B$&	$C$&	$D$&	$\sigma$\\
	\hline
	& kJ/mol&$\quad$Å$^{-1}$&	$\ $Å$^6$\,kJ/mol&$\ $Å$^8$\,kJ/mol&$\qquad$Å\\
	\hline
	Na--Na&	25.444&	3.155&	101.17&	48.18	 &       2.340\\
	Na--Cl&	20.355&	3.155&	674.48&	837.08	 &       2.755\\
	Cl--Cl&	15.266&	3.061&	7241.65&14543.47 &       3.170\\
	K--K&    25.444&	2.967&	1463.38&1445.31	 &       2.926\\
	K--Na&   25.444&	3.061&	377.31&	260.28	 &       2.633\\
	K--Cl&   20.355&	2.967&	2890.63&4396.16	 &       3.048\\
	\hline \hline
\end{tabular}
\caption{Force field parameters for NaCl--KCl.}
\label{tab:parameters} 
\end{table}

\subsection{Semi-grand canonical ensemble simulation}
\label{sec:semi-grand}

Fluid-fluid phase equilibria require thermal, mechanical, and compositional equilibrium, i.e. $T = T_1 = T_2$, $p = p_1 = p_2$, and $\mu(x_1, T, p) = \mu(x_2, T, p)$. If one can guess the approximate fluid densities and a composition between $x_1$ and $x_2$, then the equilibrium conditions can often be identified from a simple direct coexistence simulation. A system held at $T$, $p$, and intermediate overall composition will spontaneously split into two phases, one at composition $x_1$ separated by an interface from another at composition $x_2$. In a long thin simulation box with fixed cross section and pressure applied from the ends, the interface spontaneously forms perpendicular to the long axis such that surface tension exerts no pressure on the adjacent “bulk” phases. If the box is much longer than the interface thickness, the equilibrium compositions can then be estimated from the simulated concentration profiles.\cite{Alejandre1995,Rivera2003,Liu2013} Similar direct coexistence simulations have been used to estimate solubilities of certain solids.\cite{Knott2012,Manzanilla2015,Kolafa2016,Espinosa2016} An alternative collection of indirect Monte Carlo techniques, which have found widespread application, can identify phase coexistence conditions without ever simulating the interface between phases or guessing approximate densities and compositions. These include grand canonical ensemble simulations,\cite{NormanFilinov1969,frenkel_smit} the Gibbs ensemble simulations,\cite{Panagiotopoulos87} Gibbs-Duhem integration,\cite{Metha_Kofke1994,Hitchcock_Hall1999} density of states methods,\cite{Mastny_dePablo2005,Boothroyd2018} and osmotic ensemble Monte Carlo simulations.\cite{Moucka2011}

For multicomponent solid-solid equilibria, the same conditions apply at equilibrium. However, additional difficulties require methods beyond those used for fluid-fluid and fluid-solid equilibria. First, a dense crystalline solid with no vacancies has no free space for inserting or growing new particles, thus preventing the use of grand canonical simulations, osmotic ensemble simulations, and Gibbs-ensemble simulations with particle insertions. Second, a simulated periodic crystal has allowed volumes that are effectively discretized by the need to complete integer layers of the crystal, thus also preventing the use of Gibbs ensemble simulations with particle exchanges. Third, two solids in direct contact with each other exert stresses on the neighboring phase,\cite{Cahn1989} especially when they share a coherent interface as expected for KCl inclusions in NaCl or NaCl inclusions in KCl. These stresses cause long range strain (lattice distortion) in the two solids. The stresses and strains can alter equilibrium compositions,\cite{Cahn1982} so direct solid-solid coexistence simulations are also not an option.

To enable the calculation of multicomponent solid-solid equilibria, we explore the semi-grand approach by Kofke and Glandt,\cite{kofke1988} where particles are transmutated into  alternative species instead of being inserted. The acceptance probability depends on the chemical potential {\it difference} between the species, rather than the individual chemical potentials. For the binary solid solution, the semi-grand approach therefore yields the required composition ratio as function of defined chemical potential difference $\Delta \mu = \mu_B - \mu_A$.

We utilised the semi-grand ensemble for the NaCl--KCl solid solution to determine co-existence compositions at a number of temperatures, but found this strategy cumbersome and inefficient compared with the Widom's semi-grand approach detailed in Sec.\,\ref{sec:semigrandwidom}.

\subsection{Semi-grand canonical Widom simulation}

\label{sec:semigrandwidom}

The test particle method originally introduced by Widom\cite{Widom1963}
directly probes the excess chemical potential of a substance
in a simulation. In the case of a system of $N$ identical particles
of mass $m$ at pressure $p$ and temperature $T$ the isothermal-isobaric
partition function is 
\begin{align}
Z_{NpT} & \equiv\frac{\beta p}{h^{3N}}\frac{1}{N!}\int_{0}^{\infty}dV\,e^{-\beta pV}\nonumber \\
 & \int dr^{N}dp^{N}e^{-\beta H(r^{N},p^{N})},
\end{align}
where $H(r^{N},p^{N})$ is the system's Hamiltonian, $\beta=1/k_{\text{B}}T$,
$k_{\text{B}}$ is the Boltzmann constant, and $h$ is Planck's constant.
The chemical potential $\mu$ can be calculated from 
\begin{equation}
\beta\mu=\frac{\partial}{\partial N}\left[-\ln Z_{NpT}\right].\label{eq:mu_definition}
\end{equation}
The Hamiltonian is 
\begin{equation}
H(r^{N},p^{N})=\sum_{i=1}^{N}\frac{p_{i}^{2}}{2m}+U\left(r^{N}\right),
\end{equation}
so we can split $\mu$ into an ideal and excess part, 
\begin{equation}
\mu=\mu^{\text{id}}+\mu^{\text{ex}}.
\end{equation}
The ideal part is given by setting $U(r^{N})=0$, explicitly performing
the momentum integrals, and then applying Eq.\,\eqref{eq:mu_definition},
\begin{equation}
e^{-\beta\mu^{\text{id}}}\equiv\frac{1}{\beta p\Lambda^{3}},
\end{equation}
where $\Lambda=\sqrt{\beta h^{2}/2\pi m}$ is the thermal de Broglie
wavelength.

For large $N$, the derivative in Eq.\,\eqref{eq:mu_definition}
can be substituted by a finite difference $\Delta N=1$, and the logarithm
turns a difference into a ratio of partition functions. One can then
show that in the thermodynamic limit we have for the excess part 
\begin{equation}
e^{-\beta\mu^{\text{ex}}}=\frac{\beta p\left\langle Ve^{-\beta\Delta U(r_{0}|r_{1}\dots,r_{N})}\right\rangle _{N}}{N+1}.\label{eq:mu_ex}
\end{equation}
The average is taken over a uniformly distributed test particle (index~0)
in a regular $NpT$ simulation box and the factor in the exponential
is the energy change associated with the virtual insertion of this
test particle. In the case of an $NVT$ ensemble, Widom's formula
reduces to 
\begin{equation}
e^{-\beta\mu^{\text{ex}}}=\left\langle e^{-\beta\Delta U(r_{0}|r_{1}\dots,r_{N})}\right\rangle _{N}.\label{eq:mu_ex_nvt}
\end{equation}
While in principle exact, Widom's method in practice suffers from
poor convergence in dense systems or crystals.

For multi-component systems, an alternative approach to virtual particle
insertions is to average over virtual particle transmutations, as
shown by Sindzingre et al.\cite{Sindzingre1987} This makes the method
applicable to crystals, enabling the determination of the chemical
potential difference for the transmutation of one chemical species
to another in the lattice. Consider a two-component system (comprised of
components $A$ and $B$)
at temperature $T$ and pressure $p$. The total particle number is
$N=N_{A}+N_{B}$. Now, the partition function is
\begin{equation}
Z_{N_{A},N_{B},p,T}=\beta p\int_{0}^{\infty}dV\,e^{-\beta pV}\,Z_{N_{A},N_{B},V,T},
\end{equation}
where 
\begin{equation}
Z_{N_{A},N_{B},V,T}=\frac{1}{h^{3N}}\frac{1}{N_{A}!N_{B}!}\int dr^{N}dp^{N}\,e^{-\beta H(r^{N},p^{N})}
\end{equation}
and the ideal part is given by 
\begin{equation}
Z_{N_{A},N_{B},p,T}^{\text{id}}=\frac{N!}{N_{A}!N_{B}!}\frac{1}{(\beta p)^{N}}\frac{1}{\Lambda_{A}^{3N_{A}}}\frac{1}{\Lambda_{B}^{3N_{B}}},
\end{equation}
where $\Lambda_{A}=\sqrt{\beta h^{2}/2\pi m_{A}}$ is the thermal
de Broglie wavelength for species $A$ with mass $m_{A}$ and $\Lambda_{B}$
is defined accordingly. The chemical potential for species $A$ is
the derivative 
\begin{equation}
\beta\mu_{A}=-\frac{\partial}{\partial N_{A}}\ln Z_{N_{A},N_{B},p,T}=\beta(\mu_{A}^{\text{id}}+\mu_{A}^{\text{ex}}),\label{eq:betamu}
\end{equation}
with $\mu_{B}$ defined accordingly. 

Consider the difference in chemical potential, 
\begin{equation}
\Delta\mu\equiv\mu_{B}-\mu_{A}.
\end{equation}
The ideal component of the chemical potential difference can be calculated analytically,
\begin{equation}
\Delta\mu^{\text{id}}\equiv\mu_{B}^{\text{id}}-\mu_{A}^{\text{id}}=-\frac{1}{\beta}\left[\frac{3}{2}\ln\left(\frac{m_{B}}{m_{A}}\right)+\ln\left(\frac{N_{A}}{N_{B}}\right)\right].\label{eq:dmu_id}
\end{equation}
For the excess chemical potential difference, we consider the case
when a particle of type $A$ is converted into type $B$, or vice
versa. The difference in excess chemical potential is then directly
related to the exponential average of the energy change for the transmutation:
\begin{align}
\Delta\mu^{\text{ex}}\equiv\mu_{B}^{\text{ex}}-\mu_{A}^{\text{ex}} & =-\frac{1}{\beta}\ln\left\langle e^{-\beta\Delta U(A^{-},B^{+})}\right\rangle _{N_{A},N_{B}}\label{eq:dmu_formula_a}\\
 & =\frac{1}{\beta}\ln\left\langle e^{-\beta\Delta U(A^{+},B^{-})}\right\rangle _{N_{A},N_{B}}.\label{eq:dmu_formula_b}
\end{align}
This quantity can be probed using virtual Monte Carlo transmutation
moves in an otherwise standard $NpT$ simulation, which itself can be either
Monte Carlo or molecular dynamics. It is worth noting
that here, the system composition is fixed at ($N_{A}$, $N_{B}$)
while the excess chemical potential difference is the dependent variable.
This is in stark contrast to a simulation in the semi-grand canonical
ensemble, where one fixes the chemical potential difference and obtains
the composition as a result. Note Widom's semi-grand method was used in an earlier study to predict the phase diagram of MgO-MnO solid solution,\cite{Allan2001} but that work refers to the method as a semi-grand canonical simulation.

\subsection{Gibbs free energy from the chemical potential difference}

\label{sec:gibbscalculation}

In a binary mixture of species $A$ and $B$, the full Gibbs free
energy {\it per particle} is given as 
\begin{equation}
\hat{G}(N_{A},N_{B},p,T)=x_{A}\mu_{A}+x_{B}\mu_{B},
\end{equation}
where $x_{i}=N_{i}/N$ is the mole fraction of species $i$, and $\hat{G}=G/N$.
Since $x_{B}=1-x_{A}$, we have 
\begin{equation}
\frac{\partial\hat{G}}{\partial x_{B}}=\mu_{B}-\mu_{A}=\Delta\mu,\label{eq:delta_mu_derivative}
\end{equation}
which relates the Gibbs free energy to the chemical potential difference
that we obtain from simulations. Using the simplified notation $x\coloneqq x_{B}$,
one has 
\begin{equation}
\hat{G}(x)=\hat{G}_{A}+\int_{0}^{x}dx'\,\Delta\mu(x').\label{eq:gbar_int}
\end{equation}
In other words, but for the reference value $\hat{G}_{A}$ (the Gibbs
free energy of a pure $A$ system), the calculation of $\Delta\mu$
as a function of system composition is sufficient to construct its
Gibbs free energy as a function of composition, enabling the determination
of co-existence points and hence the binary phase diagram. 

In the context of mixtures, the Gibbs free energy is often written
in a slightly different form, as 
\begin{equation}
\hat{G}=\hat{G}^{I}+\hat{G}^{E}.\label{eq:g_total}
\end{equation}
Again, $\hat{G}$ is split into an ideal and excess part, but “ideal”
and “excess” in this case have a different meaning. The ideal Gibbs
energy is now given as 
\begin{equation}
\frac{\hat{G}^{I}(x)}{k_{\text{B}}T}=(1-x)\frac{\hat{G}_{A}}{k_{\text{B}}T}+x\frac{\hat{G}_{B}}{k_{\text{B}}T}+(1-x)\ln(1-x)+x\ln x,\label{eq:g_ideal}
\end{equation}
where $\hat{G}_{A}$ and $\hat{G}_{B}$ describe the pure end states
(including all non-idealities in the sense of Sec.\,\ref{sec:semigrandwidom}).
The ideal Gibbs energy of the mixture is a composition-weighted linear
combination of the pure states plus two mixing entropy terms. This
result is consistent with explicitly integrating the ideal gas case,
i.\,e. using Eq.\,\eqref{eq:dmu_id} in Eq.\,\eqref{eq:gbar_int},
and noting that in the special case of two ideal gases that only differ
in particle mass, we have 
\begin{equation}
\frac{\hat{G}_{B}}{k_{\text{B}}T}=\frac{\hat{G}_{A}}{k_{\text{B}}T}-\frac{3}{2}\ln\frac{m_{B}}{m_{A}}.
\end{equation}
Naturally, for any system with non-zero interactions (such as NaCl--KCl),
this simple relation between $\hat{G}_{A}$ and $\hat{G}_{B}$ does
not hold true.

The change of notation is intended here as we want to emphasize that
we will in the following assume some functional form for the excess
part and fit it to the simulation results. The simplest choice for
$G^{E}$ is called the regular solution model, and is usually written
in the form 
\begin{equation}
\frac{G^{E}}{k_{\text{B}}T(1-x)x}=\Omega.
\end{equation}
The regular solution model is symmetric with respect to $A$--$B$
interactions, i.\,e. $A$ in a $B$ solvent has the same $G^{E}$
as $B$ in an $A$ solvent. The single parameter $\Omega$ measures
the difference in interaction between the two species. More explicitly,
\begin{equation}
\Omega=N_{b}\left(\varepsilon_{AB}-\frac{\varepsilon_{AA}+\varepsilon_{BB}}{2}\right),
\end{equation}
where $N_{b}$ is the number of bonds per particle and $\varepsilon_{ij}$
is the interaction energy of a bond between species $i$ and $j$.
Our results show that a regular solution model is not sufficient
for the NaCl--KCl mixture. Asymmetry can be introduced with the two-parameter
Margules model, 
\begin{equation}
\frac{G^{E}}{k_{\text{B}}T(1-x)x}=W_{1}(1-x)+W_{2}x,\label{eq:g_margules}
\end{equation}
which we use to fit our simulation data. Specifically, we combine
Eqs.\,\eqref{eq:g_total}, \eqref{eq:g_ideal}, and \eqref{eq:g_margules}
to get the full Gibbs free energy and then take the derivative with
respect to $x$ in order to obtain $\Delta\mu$, Eq.\,\eqref{eq:delta_mu_derivative}.
The final functional form we use to fit $\Delta\mu(x)$ is hence 
\begin{equation}
\frac{\Delta\mu}{k_{\text{B}}T}=W_{0}+\ln\left(\frac{x}{1-x}\right)+W_{1}(1-x)(1-3x)-W_{2}x[1-3(1-x)],\label{eq:dmu_fit_form}
\end{equation}
where the parameter $W_{0}=(\hat{G}_{B}-\hat{G}_{A})/k_{\text{B}}T$
can be identified as the chemical potential difference of the pure
phases. The corresponding Gibbs free energy per particle is 
\begin{align}
\frac{\hat{G}(x)}{k_{\text{B}}T}= & xW_{0}+(1-x)\ln(1-x)+x\ln(x)\nonumber \\
 & +x(1-x)[W_{1}(1-x)+W_{2}x].
\end{align}
Here, we have used the normalization $\hat{G}_{A}=0$. As we will
see momentarily, knowledge of the absolute value is not important
for the construction of the phase diagram. However, should the Gibbs
free energies $\hat{G}_{A}$ and $\hat{G}_{B}$ of the pure end states
be known, the fitted value of $W_{0}$ can serve as a consistency
check.

\subsection{Construction of the phase diagram}

\label{sec:phasediagcalculation}

\begin{figure}
\centering \includegraphics[width=1\columnwidth]{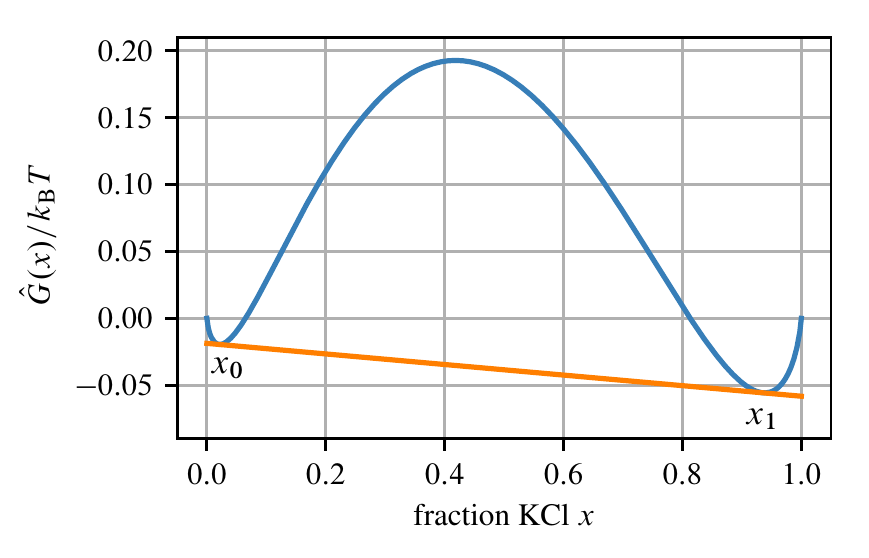}
\caption{(Blue) Gibbs free energy as a function of composition when there is
two-phase coexistence. The data shown are for the NaCl--KCl system
at a temperature of 500\,K, with $x$ denoting the mole fraction of KCl.
Note that the linear part of the Gibbs energy has been subtracted,
as this will not affect the final result. (Orange) The double tangent
construction yields the coexistence concentrations $x_{0}$ and $x_{1}$.}
\label{fig:double_tangent_margules} 
\end{figure}

Consider the Gibbs free energy as a function of system composition
as shown in Fig.\,\ref{fig:double_tangent_margules} (blue curve).
Two-phase coexistence occurs if this function has a concave region.
A homogeneous system with a concentration $x_{0}<x<x_{1}$ is not
thermodynamically stable, and the two coexistence concentrations $x_{0}$
and $x_{1}$ are found with a double tangent construction in the diagram.
There, a homogeneous system can lower its free energy by going down
to the orange line, the double tangent. Physically, the linear function
corresponds with a system that consists of two separated phases of
composition $x_{0}$ ($A$-rich) and $x_{1}$ ($B$-rich), ignoring
any interface terms. The double tangent to a function $f(x)$ is found
by solving a system of two equations with two unknowns, 
\begin{align}
f'(x_{0}) & =f'(x_{1}),\\
f(x_{0})+f'(x_{0})(x_{1}-x_{0}) & =f(x_{1}).
\end{align}
An important thing to note is that one can replace $f(x)$ by $g(x)=f(x)+kx$,
i. e. add an arbitrary linear term, without changing these equations
and hence the result for $(x_{0},x_{1})$. In practice, almost always
the equations have to be solved numerically. To give an example, even
in the simple case of a regular solution model, the resulting equation
\begin{equation}
\ln\frac{x}{1-x}=\Omega(2x-1),
\end{equation}
does not have an algebraic solution.

\subsection{Simulation details}

The primary simulations to obtain the phase diagram were carried out
using an in-house code, ATOMH.\cite{atomh} We also used LAMMPS\cite{lammps}
and DL\_POLY\cite{Todorov2006} for auxiliary calculations. If not
otherwise stated, the system size was $N=256$ ion pairs, or $512$
particles, and the pressure $p=1$\,bar. The cutoff for the non-bonded
van der Waals and real-space Coulombic interactions was 0.9 nm. Long range
van der Waals corrections were applied to the energy and the virial. The
long-range part of the Coulomb interaction was calculated using Ewald
summation, with the relative accuracy parameter set to $10^{-6}$.
For MD simulations, the time step was $\Delta t=2$\,fs. For MC simulations,
one sweep corresponded to 512 single-particle moves, and on average 50 exchange moves
(Na and K swaps) and 10 volume moves. The quantity of interest, the
chemical potential difference $\Delta\mu,$ was sampled every 50 sweeps
by performing a virtual transmutation of every Na and K atom to the
opposite type and updating the corresponding average, Eqs.\,\eqref{eq:dmu_formula_a}
and \eqref{eq:dmu_formula_b}.

\section{Results and discussion}

\label{sec:results_discussion}

\subsection{Force field checks}

For a first check of the force field accuracy, we performed
simulations of the pure NaCl and KCl crystalline phases, respectively.
The average lattice energies at $T=298$\,K were $U$(NaCl)
= $-769.95\pm 0.01$ kJ mol$^{-1}$ per ion-pair, and $U$(KCl) = $-700.43\pm
0.01$ kJ mol$^{-1}$ per ion-pair, which are in excellent agreement with experimentally determined lattice energies,
$U_{\textrm{expt}}$(NaCl) = $-770.3$ kJ mol$^{-1}$ per ion-pair, and $U_{\textrm{expt}}$(KCl) = $-701.2$ kJ mol$^{-1}$ per ion-pair.\cite{Datta2012}

\begin{figure}
\centering \includegraphics[width=1\columnwidth]{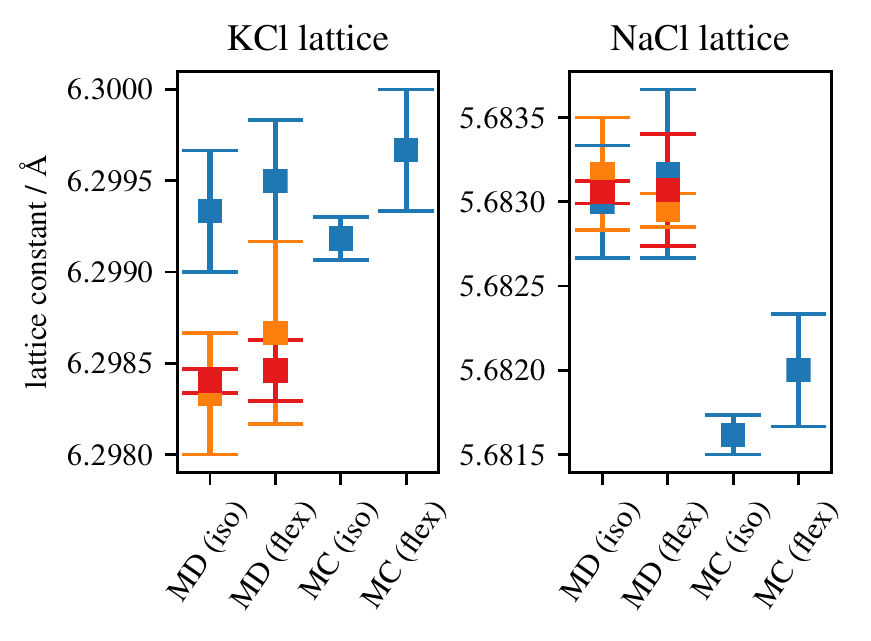}
\caption{Lattice constants for the pure phases and estimated uncertainties at 298\,K. (Blue) ATOMH code results. (Orange) DL\_POLY code results. (Red) LAMMPS code results. Iso denotes an isotropic barostat implying a cubic simulation box, and flex a fully flexible box. System size is $N=864$ ion pairs (1728 particles, 6~unit cells in each direction).}
\label{fig:lattice_constants}
\end{figure}

\begin{figure}
\centering \includegraphics[width=1\columnwidth]{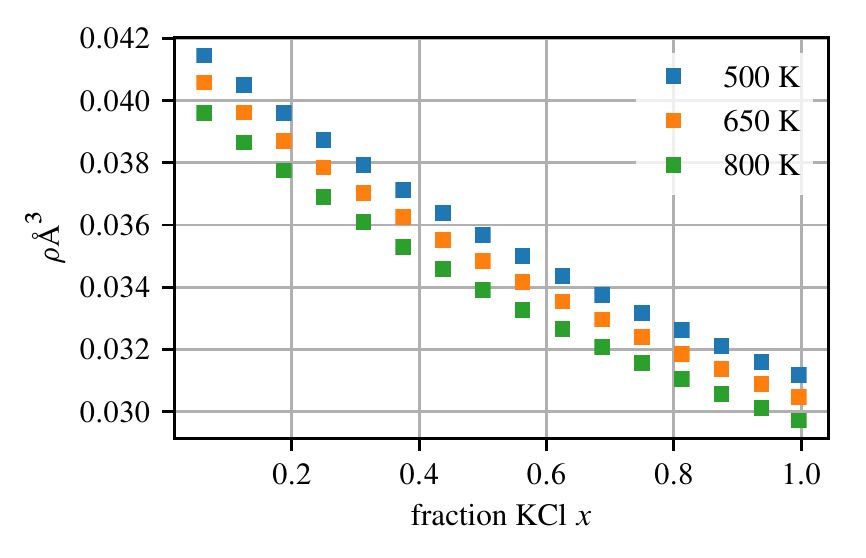}
\caption{Densities as a function of system composition at three selected temperatures.}
\label{fig:densities} 
\end{figure}

The results for the lattice constants at $T=298$\,K are given in Fig.\,\ref{fig:lattice_constants}.
Accounting for the estimated uncertainty, there is good agreement
between both the MC and MD results as well as different simulation
packages. There are eight atoms per unit cell, so lattice constant and
(number) density are related by $\rho = 8 / a^3$. The 
densities are $\rho(\text{NaCl}) = 0.043595 \pm 0.000007$\,Å$^{-3}$
and $\rho(\text{KCl}) = 0.032008 \pm 0.000004$\,Å$^{-3}$. The corresponding
mass densities are $\rho_m(\text{NaCl}) = 2.1154 \pm 0.0004$\,g/cm$^{3}$
and $\rho_m(\text{KCl}) = 1.9812 \pm 0.0002$\,g/cm$^{3}$. Densities of the mixture as a function of composition for a few other selected temperatures are shown in Fig.\,\ref{fig:densities}. In the range of temperatures investigated, there is an approximately linear relationship similar to Vegard's law\cite{Vegard1921} between lattice constant, composition, and temperature:
\begin{equation}
a \approx c_x x + c_T T + a_0.
\end{equation}

The fitted parameter values are $c_x = 0.62796$\,Å, $c_T = 3.1485 \times 10^{-4}$\,Å/K, and $a_0 = 5.5911$\,Å.

\subsection{Free energies of the pure end states}

We calculated the (Helmholtz) free energies of pure NaCl ($x=0$)
and pure KCl ($x=1$) using the Einstein crystal approach via thermodynamic
integration (TI). The calculations closely followed those in a previous
work of Anwar, Frenkel, and Noro,\cite{Anwar2003} so only a brief
overview is given here. The simulations were carried out at a fixed
volume corresponding to a pressure of $p=1\,$bar. We used Gauss--Legendre
quadrature with $n=16$ $\lambda$ nodes for the numerical integration
of the free energy derivative $\partial F / \partial\lambda$.
The equations are provided in Supporting Information. In the Einstein crystal part of the calculation, the thermal de
Broglie wavelength $\Lambda$ was set to 1\,Å for all species. For the conditions investigated
here, the $pV$ term to convert the Helmholtz free energy into a Gibbs
free energy is very small and hence deemed unimportant. For example,
for NaCl at 298\,K, the correction is $pV_{\text{NaCl}}/N_{\text{NaCl}}k_{\text{B}}T\approx0.001$,
much smaller than even the probably slightly underestimated uncertainty
in the Helmholtz free energy. The results are summarized in Tab.\,\ref{tab:free_energies}.

\begin{table}[htbp]
\centering 
\global\long\def\arraystretch{1.5}%
\begin{tabular}{l|r r r r r}
	\hline \hline
	$T$ & $\hat{F}_{\text{NaCl}} / k_{\text{B}} T$ &	$\hat{F}_{\text{KCl}} / k_{\text{B}} T$ \\
	\hline
	298 & $-306.525 \pm 0.005$ & $-279.119 \pm 0.004$ \\
	600 & $-150.766 \pm 0.005$ & $-137.474 \pm 0.005$ \\
	\hline \hline
\end{tabular}
\caption{Helmholtz free energies for pure crystalline phases of NaCl and KCl from thermodynamic
integration with the thermal de Broglie wavelength set to 1\,Å for all species to enable easier comparison with published values of Aragones, Sanz, and Vega.\cite{Aragones2012} In analogy to $\hat{G}$, $\hat{F}$ denotes the Helmholtz
free energy per ion pair.}
\label{tab:free_energies}
\end{table}

For NaCl at 298\,K, our result is $\hat{F}_{\text{NaCl}} / k_{\text{B}} T = -306.525 \pm 0.005$. The result compares well to that of Aragones, Sanz, and Vega,\cite{Aragones2012} which is $-306.22\,k_{\text{B}} T$. Note that these authors report the value \textit{per particle}, while in our study all values are given \textit{per ion pair}, which accounts for the factor two difference to the raw result reported in Tab.\,II of the reference. As discussed in Sec.\,\ref{sec:model}, the Cl--Cl $B$, $C$, and $D$ potential parameters utilized in the present study differ from the Cl--Cl parameters of pure NaCl in the study by Aragones, Sanz, and Vega. Our $B$, $C$, and $D$ parameters are averages from the Cl--Cl parameters for pure NaCl and KCl systems, possibly contributing to the 0.3\,$k_{\text{B}} T$ difference between our NaCl free energies and theirs.

\subsection{Phase diagram calculation}

\begin{figure}
	\centering \includegraphics[width=1\columnwidth]{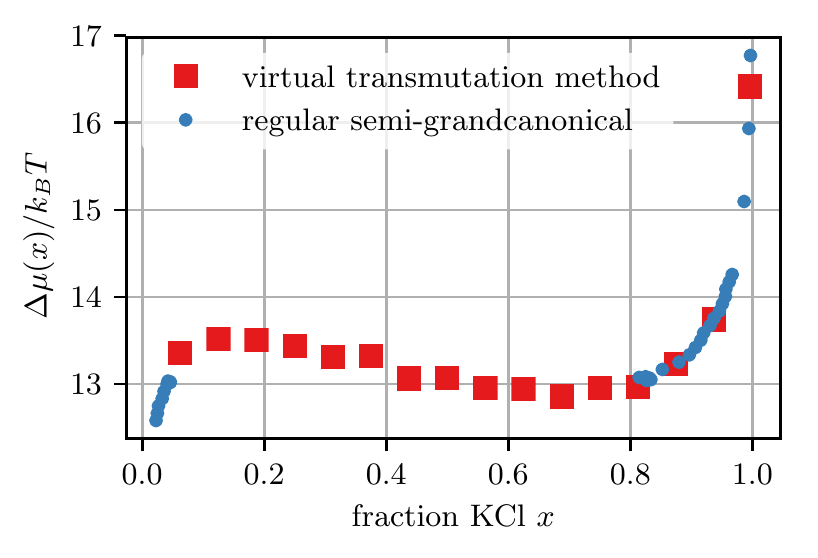}
	\caption{Comparison between a regular semi-grand canonical simulation and the
		Widom-like virtual transmutation method for NaCl--KCl at $T=600$\,K. It is not possible to access the miscibility gap region with the regular semi-grand method.}
	\label{fig:semigrand-comparison} 
\end{figure}

The miscibility gap is theoretically unstable and inaccessible to regular semigrand canonical
simulations, an issue which was noted by Sadigh et\,al.\cite{Sadigh2012} Given a
fixed $\Delta\mu$, the semigrand canonical simulations always find a composition on either
side of the miscibility gap and stay there. Without a good initial guess for the
value of $\Delta\mu$ at coexistence, one needs to
find it in a tedious, iterative procedure. For example, one can run semigrand simulations with a coarse grid of $\Delta\mu$ values and then iteratively improve the resolution of $\Delta\mu$ at coexistence.  Results from this procedure at $T=600$\,K are shown in blue in Fig.\,\ref{fig:semigrand-comparison}. This procedure must be repeated for each temperature below the critical point.  

Fig.\,\ref{fig:semigrand-comparison} also shows results from Widom's semi-grand method. While both methods agree on the overall curve, only Widom's semi-grand method is able to probe the miscibility
gap. Furthermore, for the Widom method, the total number of simulations necessary to run
is much lower, as the whole curve can be obtained in a single parameter
sweep varying the KCl mole fraction from 0 to 1 in equal increments.

For the Widom method, an NaCl--KCl solid solution of any arbitrary composition is unable to undergo phase separation into the two co-existing phases due to interfacial energy barriers, enabling its chemical potential to be monitored. This opportunity of being able to sample the chemical potential of unstable and metastable structures is essential to the application of the double tangent method. So, it appears that the same interfacial barriers that limit the study of phase transition phenomenon by brute force simulation serve here to make the miscibility gap accessible. To confirm that the system does not phase-separate
into an Na-rich and a K-rich domain over the full length of our simulations, we carried out spatial and temporal correlation analysis. The analysis and associated results are detailed in Supporting Information and indeed confirm the systems remain homogeneous.  For larger systems which do phase-separate, one would need to employ alternative sampling techniques to probe the miscibility gap, such as the generalized replica exchange method developed by Kim, Keyes, and Straub.\cite{Kim2010}

Our experience of Widom's semi-grand approach suggests that this methodology would readily carry over to more complex solid solution systems. There appears to be a remarkable coupling of this method with solid solution phase stability. Convergence of the method requires appreciable overlap of the energy distributions of the original system and that of the virtual perturbed state (when a molecule has been transmutated). This translates to the transmutation energy differences not being excessively large. Phase stability of a solid solution relies on exactly the same energy criteria---the substitution energy difference again must not be too large. The implication is that any system that forms a solid solution is likely to be accessible to Widom's semi-grand sampling, which would include flexible molecular systems too.

\begin{figure}
\centering \includegraphics[width=1\columnwidth]{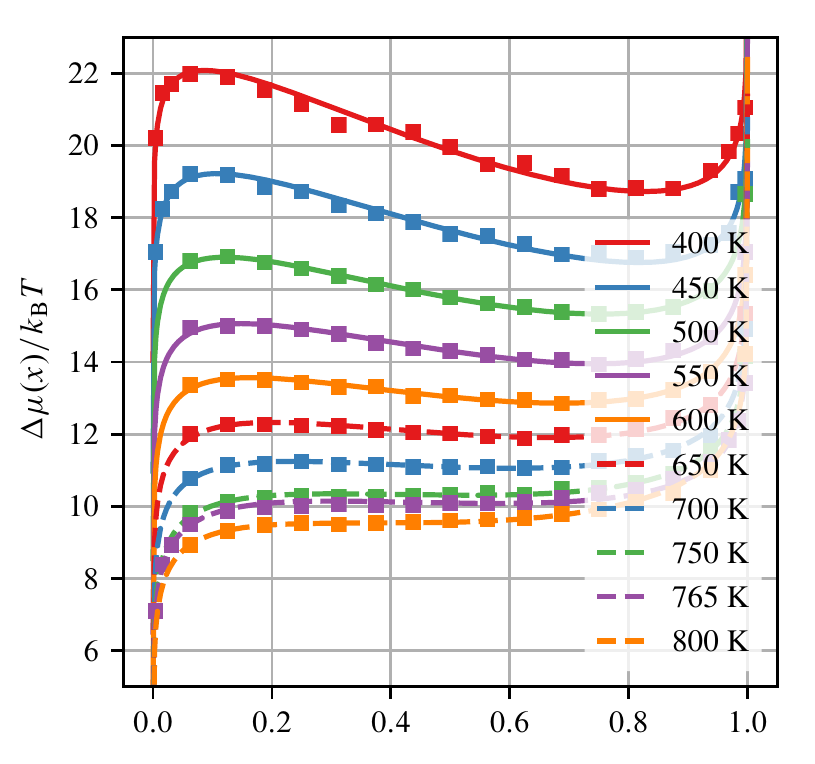}
\caption{Chemical potential difference for NaCl--KCl at a range of temperatures.
$x$ is the fraction of KCl. Each square is the result of a simulation
at a fixed composition. The lines are fits with Eq.\,\eqref{eq:dmu_fit_form}.}
. \label{fig:dmu_alltemps} 
\end{figure}

We have performed simulations covering the full range of compositions
for multiple temperatures (Fig.\,\ref{fig:dmu_alltemps}). The obtained
fit parameters $W_{1}$ and $W_{2}$ are used to find the corresponding
coexistence concentrations $x_{0}$ and $x_{1}$ via a double tangent
construction. The highest temperature, 800\,K, is above the critical
point. This is already obvious from the functional form of $\Delta\mu(x)$,
which is strictly monotonic. Consequently, the corresponding Gibbs
free energy is convex, and no phase coexistence occurs. The results
for $(x_{0},x_{1})$ are shown in Fig.\,\ref{fig:ftsalt} and plotted
on top of the experimental phase diagram from the literature. Note that in this
figure, the x~axis denotes the mole fraction of NaCl. The accuracy of the prediction is remarkable, highlighting the quality of the NaCl--KCl forcefield, validating Thakkar's mixing rule,\cite{Thakkar1984} and the power of Widom's semi-grand simulation protocol. The largest deviations between the experimental results and the ones computed in this study appear at higher temperatures near the critical point. Presumably, this is due to larger fluctuations in the vicinity of the critical point, which make it harder for the computational results to converge.

\begin{figure}
\centering \includegraphics[width=1\columnwidth]{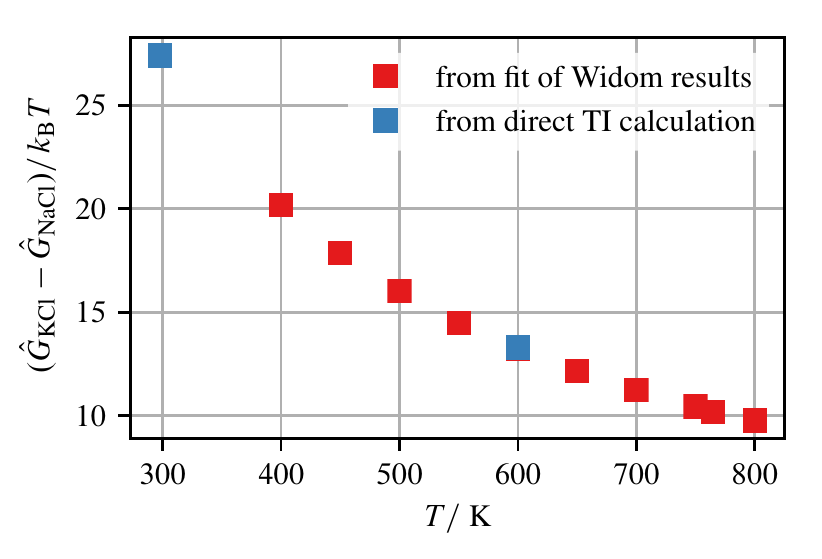}
\caption{Gibbs free energy difference between the pure phases obtained from
the fitting constant $W_{0}$ and from independent Einstein crystal-based calculations for the pure phases. For
$T=298\,$K, the two results are so close that only the topmost marker
(blue) is visible.}
\label{fig:consistency_check} 
\end{figure}

To verify the consistency of our simulations, 
we show the results for the Gibbs free energy difference between the
two pure phases ($\hat{G}_{\textrm{KCl}}-\hat{G}_{\textrm{NaCl}}$) as a function of temperature in Fig.\,\ref{fig:consistency_check}. The data shown here is derived from both the independently determined free energies of the pure end states and from the semi-grand canonical Widom simulations. For the Widom simulations the free energy difference corresponds directly
to the value of the fit parameter $W_{0}$. The free energies of the pure KCl and NaCl are those
calculated using thermodynamic integration starting from an Einstein crystal at $T=298\,$K and $T=600\,$K. As seen in the figure, we have almost perfect
agreement between the two methods. Note that the semi-grand canonical
Widom results for $T=298\,$K are not shown in Fig.\,\ref{fig:dmu_alltemps},
we use them only for the consistency check.

\subsection{Critical point extrapolation}

\begin{figure}
\centering \includegraphics[width=1\columnwidth]{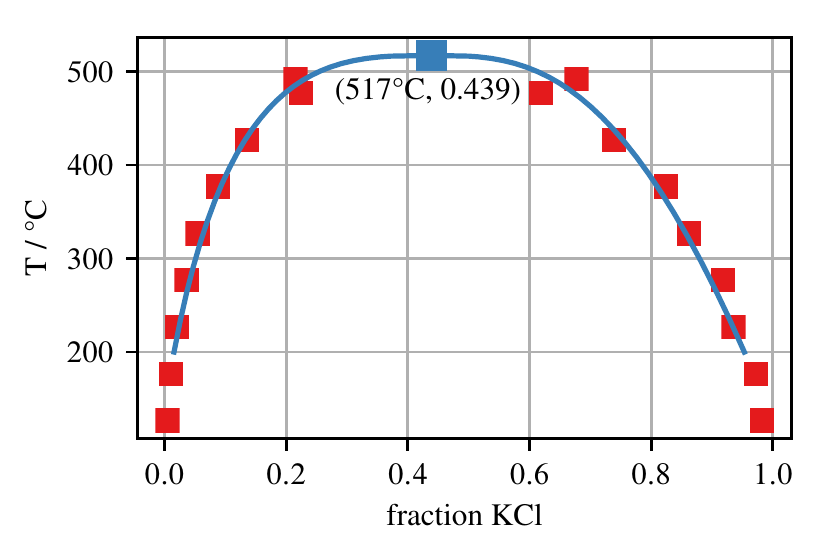}
\caption{Extrapolated critical point for NaCl--KCl using the Guggenheim formula,
Eqs.\,\eqref{eq:guggenheim_1} and \eqref{eq:guggenheim_2}. The red data points are the same as shown in Fig.\,\ref{fig:ftsalt}, but note that the x-axis is reversed compared to Fig.\,\ref{fig:ftsalt}.}
\label{fig:critical_point}
\end{figure}

We use Guggenheim's empirical formula\cite{Guggenheim1972} to extrapolate
the critical point. The original formula is for the critical point
in a gas-liquid transition of argon, and it gives the gas ($\rho_{g}$)
and liquid ($\rho_{l}$) densities as a function of temperature: 
\begin{align}
\frac{\rho_{l}}{\rho_{c}}=1+0.75\left(\frac{T_{c}-T}{T_{c}}\right)+1.75\left(\frac{T_{c}-T}{T_{c}}\right)^{1/3},\label{eq:guggenheim_1}\\
\frac{\rho_{g}}{\rho_{c}}=1+0.75\left(\frac{T_{c}-T}{T_{c}}\right)-1.75\left(\frac{T_{c}-T}{T_{c}}\right)^{1/3}.\label{eq:guggenheim_2}
\end{align}
The critical point is $(T_{c},\rho_{c})$. In the present application,
mole fraction $x$ is substituted for the density. We allow both
the critical point as well as the numerical parameters 0.75 and 1.75
to change in order to give the best fit to our data. The result is
shown in Fig.\,\ref{fig:critical_point}.

\section{Concluding remarks}

\label{sec:conclusions}

We have demonstrated that the semi-grand canonical Widom method\cite{Sindzingre1987} can accurately and efficiently compute the phase diagram for a multicomponent solid as a function of temperature and composition. Specifically, we have computed the binodal for solid solutions of Tosi--Fumi NaCl and KCl below the critical temperature. We used the semi-grand Widom framework to compute the chemical potential difference between NaCl and KCl components as a function of composition. The chemical potential data were converted to ideal and excess free energies, fitted to a Margules model, and then used to determine coexistence compositions by common tangent constructions.

Our results are in remarkable agreement with the phase diagram based
on experimental data, a testament to the Fumi-Tosi force field, Thakkar's mixing rule, and the robust convergence of the Widom's semi-grand method.   Perhaps more importantly, our results provide
coexistence conditions and chemical potentials at metastable compositions
within the binodal. Thus the results open the door to quantitative studies of nucleation, growth, and coarsening processes within solid solutions.

\section{Supplementary Material}

In the Supplementary Material, we provide details on the thermodynamic integration procedure to calculate the free energies of the pure salts and provide absolute free energy results using the true thermal de Broglie wavelengths. We also compare our version of the Cl--Cl interaction with the one used in earlier studies and show the lattice parameter of the mixture for different temperatures. Furthermore, we perform a spatial and temporal correlation analysis to ensure compositional homogeneity throughout the unstable zone of the phase diagram.

\section{Acknowledgments}

\label{sec:ack}

BP and CL acknowledge financial support from the National Science
Foundation Award No. 1465289 in the Division of Theoretical Chemistry.
We thank Mikhail Laventiev, Kartik Kamat, and Daan Frenkel for helpful discussions.

\appendix

\end{document}